\documentclass[sigplan,10pt]{acmart}

\copyrightyear{2021}
\acmYear{2021}
\setcopyright{rightsretained}
\acmConference[PaPoC'21]{8th Workshop on Principles and Practice of Consistency for Distributed Data}{April 26, 2021}{Online, United Kingdom}
\acmBooktitle{8th Workshop on Principles and Practice of Consistency for Distributed Data (PaPoC'21), April 26, 2021, Online, United Kingdom}
\acmDOI{10.1145/3447865.3457962}
\acmISBN{978-1-4503-8338-7/21/04}

\usepackage{enumitem}
\usepackage{pervasives}
\usepackage{subcaption}

\newcommand{\library}{Quoracle}

\begin{document}

\title{Read-Write Quorum Systems Made Practical}

\author{Michael Whittaker}
\email{mjwhittaker@berkeley.edu}
\affiliation{%
  \institution{UC Berkeley}
  \city{}
  \state{}
  \country{}
}

\author{Aleksey Charapko}
\email{aleksey.charapko@unh.edu}
\affiliation{%
  \institution{University of New Hampshire}
  \city{}
  \state{}
  \country{}
}

\author{Joseph M. Hellerstein}
\email{hellerstein@berkeley.edu}
\affiliation{%
  \institution{UC Berkeley}
  \city{}
  \state{}
  \country{}
}

\author{Heidi Howard}
\email{heidi.howard@cl.cam.ac.uk}
\affiliation{%
  \institution{University of Cambridge}
  \city{}
  \state{}
  \country{}
}

\author{Ion Stoica}
\email{istoica@berkeley.edu}
\affiliation{%
  \institution{UC Berkeley}
  \city{}
  \state{}
  \country{}
}

{\begin{abstract}
Quorum systems are a powerful mechanism for ensuring the consistency of
replicated data. Production systems usually opt for majority quorums due to
their simplicity and fault tolerance, but majority quorum systems provide poor
throughput and scalability. Alternatively, researchers have invented a number
of theoretically ``optimal'' quorum systems, but the underlying theory ignores
many practical complexities such as machine heterogeneity and workload skew.
In this paper, we conduct a pragmatic re-examination of quorum systems. We
enrich the current theory on quorum systems with a number of practical
refinements to find quorum systems that provide higher throughput, lower
latency, and lower network load. We also develop a library \library{} that
precisely quantifies the available trade-offs between quorum systems to empower
engineers to find optimal quorum systems, given a set of objectives for
specific deployments and workloads. Our tool is available online at:
\url{https://github.com/mwhittaker/quoracle}.
\end{abstract}

}

\ccsdesc[500]{Computer systems organization~Reliability}
\ccsdesc[500]{Computer systems organization~Availability}
\ccsdesc[500]{Computer systems organization~Redundancy}
\ccsdesc[500]{Software and its engineering~Cloud computing}

\keywords{Quorum Systems, Read-write Quorum Systems, Distributed Systems, %
          Consensus, State Machine Replication}

\maketitle

{\section{Introduction}\seclabel{Introduction}

Ensuring the consistency of replicated data is a fundamental challenge in
distributed computing. One widely utilized solution is to require that any
operation over replicated data involve a quorum of machines. A
\defword{read-write quorum system} consists of a set of read quorums and a set
of write quorums such that every read quorum and every write quorum intersect.
Data is read from a read quorum of machines, and data is written to a write
quorum of machines.  This ensures that all previous writes are observed by
subsequent reads. In addition to data replication~\cite{Gifford79}, quorum
systems have been applied to consensus algorithms~\cite{Fischer85, Lamport98,
Howard2017, Charapko19}, distributed databases~\cite{Thomas79}, abstract data
types~\cite{Herlihy86}, mutual exclusion~\cite{Maekawa85, Agrawal89} and shared
memory~\cite{Attiya95} to name but a few.

\defword{Majority quorum systems}---quorum systems where every read and write
quorum consists of a strict majority of machines---are widely used in practice.
Their simplicity makes them well-understood, and they also tolerate an optimal
number of faults, $\left \lfloor{\frac{n-1}{2}}\right \rfloor$ with $n$
machines.  However, the performance of majority quorum systems is far from
ideal. If each machine can process $\alpha$ commands per second then the
maximum throughput of a majority quorum system is limited to just $2\alpha$,
regardless of the number of machines~\cite{Naor98}.

The academic literature has responded by proposing many quorum systems
including Crumbling Walls~\cite{Peleg95}, Trees~\cite{Agrawal90}, weighted
voting~\cite{Gifford79,Garcia-Molina85}, multi-dimensional
voting~\cite{Cheung90}, Finite Projective Planes~\cite{Maekawa85},
Hierarchies~\cite{Kumar91}, Grids~\cite{Cheung92}, and Paths~\cite{Naor98}.
These sophisticated quorum systems are rarely used for two reasons. First, the
theory behind these quorum systems ignores many practical considerations such
as machine heterogeneity, workload skew, latency, and network load. As we show
in \secref{CaseStudy}, ``theoretically optimal'' quorum systems often
underperform in practice. Second, understanding the various quorum systems and
choosing the one that is optimal for a given workload is difficult and
sensitive to workload parameters.

This paper is a practical re-examination of read-write quorum systems.  We
revisit the mathematical theory of quorum systems with a pragmatic lens and
the ambition to make less-frequently used quorum systems more broadly
accessible to the engineering community. More concretely, we make the
following contributions:
\begin{enumerate}[leftmargin=*]
  \item
    We add a number of practical refinements to the theory of read-write quorum
    systems (\S\ref{sec:Definitions}). We extend definitions to accommodate
    heterogeneous machines and shifting workloads; we introduce the notion of
    $f$-resilient strategies to make it easier to trade off performance for
    fault tolerance; and we integrate metrics of latency and network load
    (\S\ref{sec:PracticalRefinements}).

  \item
    We develop a Python library, called \library{} (Quorum Oracle), that allows
    users to model, analyze, and optimize read-write quorum systems
    (\S\ref{sec:PracticalRefinements}). We also provide a heuristic search
    procedure to find quorum systems that are optimized with respect to a
    number of user provided objectives and constraints. Given the complex
    trade-off space, we believe that using an automated assistance library like
    ours is the only realistic way to find good quorum systems.

  \item
    We perform a case study showing how to use \library{} to find quorum
    systems that provide $2\times$ higher throughput or $3 \times$ lower
    latency than naive majority quorums (\S\ref{sec:CaseStudy}).
\end{enumerate}
}
{\section{Definitions}\seclabel{Definitions}
In this section, we present definitions adapted from the existing theory on
quorum systems~\cite{Naor98, Ibaraki93}.

\subsection{Read-Write Quorum Systems}\seclabel{QuorumSystemsDefinition}
Given a set $X = \set{x_1, \ldots, x_n}$, a \defword{read-write quorum
system}~\cite{Naor98} over $X$ is a pair $Q = (R, W)$ where
\begin{enumerate}
  \item
    $R$ is a set of subsets of $X$ called \defword{read quorums},
  \item
    $W$ is a set of subsets of $X$ called \defword{write quorums}, and
  \item
    every read quorum intersects every write quorum. That is, for every $r \in
    R$ and $w \in W$, $r \cap w \neq \emptyset$.\footnote{%
      Note that some papers (e.g.~\cite{Kumar91,Cheung92}) use a more
      restrictive definition of read-write quorum systems which additionally
      requires that any two write quorums intersect.
    }
\end{enumerate}

For example, the majority quorum system over the set $X = \set{a, b, c}$ is
$Q_\text{maj} = (R, W)$ where $R = W = \set{\set{a, b}, \set{b, c}, \set{a,
c}}$.
If every read quorum intersects every write quorum, then any superset of a read
quorum intersects any superset of a write quorum. Thus, if a set $r$ is a
superset of any read quorum in $R$, we consider $r$ a read quorum as well.
Similarly, if a set $w$ is a superset of any write quorum in $W$, we consider
$w$ a write quorum. For example, we consider the set $\set{a, b, c}$ a read and
write quorum of $Q_\text{maj}$ even though the set $\set{a, b, c}$ is not
listed explicitly in $R$ or $W$.

It is notationally convenient to denote sets of read and write quorums over a
set $X$ as boolean expressions over $X$~\cite{Ibaraki93}. For example, we can
represent the set $\{\set{a, b}$, $\set{b, c}$, $\set{a, c}\}$ as the
expression $(a \land b) \lor (b \land c) \lor (a \land c)$, which we abbreviate
as $ab + bc + ac$.  Equivalently, we can express the set as $a(b + c) + bc$,
$b(a + c) + ac$, or $c(a + b) + ab$.
As another example, consider the 2 by 3 grid quorum system $Q_{2 \times 3}$
over the set $X = \set{a, b, c, d, e, f}$ as shown in \figref{TwoByThreeGrid}.
Every row is a read quorum, and every column is a write quorum. Concretely,
$Q_{2 \times 3} = \text{($abc + def$, $ad + be + cf$)}$.

\begin{figure}[ht]
  \centering

  \tikzstyle{element}=[draw, circle, inner sep=0pt, minimum width=14pt, thick]
  \tikzstyle{quorum}=[rounded corners, thick]
  \tikzstyle{readquorum}=[quorum, red]
  \tikzstyle{writequorum}=[quorum, blue]

  \begin{subfigure}[b]{0.5\columnwidth}
    \centering
    \begin{tikzpicture}
      \node[element] (a) at (0, 1) {$a$};
      \node[element] (b) at (1, 1) {$b$};
      \node[element] (c) at (2, 1) {$c$};
      \node[element] (d) at (0, 0) {$d$};
      \node[element] (e) at (1, 0) {$e$};
      \node[element] (f) at (2, 0) {$f$};
      \draw[readquorum] (-0.3, 1.3) rectangle (2.3, 0.7);
      \draw[readquorum] (-0.3, 0.3) rectangle (2.3, -0.3);
    \end{tikzpicture}
    \caption{Read quorums $abc + def$}
  \end{subfigure}%
  \begin{subfigure}[b]{0.5\columnwidth}
    \centering
    \begin{tikzpicture}
      \node[element] (a) at (0, 1) {$a$};
      \node[element] (b) at (1, 1) {$b$};
      \node[element] (c) at (2, 1) {$c$};
      \node[element] (d) at (0, 0) {$d$};
      \node[element] (e) at (1, 0) {$e$};
      \node[element] (f) at (2, 0) {$f$};
      \draw[writequorum] (-0.3, 1.3) rectangle (0.3, -0.3);
      \draw[writequorum] (0.7, 1.3) rectangle (1.3, -0.3);
      \draw[writequorum] (1.7, 1.3) rectangle (2.3, -0.3);
    \end{tikzpicture}
    \caption{Write quorums $ad + be + cf$}
  \end{subfigure}
  \caption{The 2 by 3 grid quorum system $Q_{2 \times 3}$.}
  \figlabel{TwoByThreeGrid}
\end{figure}
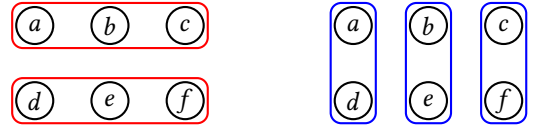

In practice, $X$ might be a set of machines, a set of locks, a set of memory
locations, and so on. In this paper, we assume that $X$ is a set of machines we
call \defword{nodes}. We assume that protocols contact a read quorum of nodes
to perform a read and contact a write quorum of nodes to perform a write.

\subsection{Fault Tolerance}
Unfortunately machines fail, and when they do, some quorums become unavailable.
For example, if node $a$ from the 2 by 3 grid quorum system $Q_{2 \times 3}$
fails, then the read quorum $\set{a, b, c}$ and the write quorum $\set{a, d}$
are unavailable.
The \defword{read fault tolerance} of a quorum system is the largest number $f$
such that despite the failure of any $f$ nodes, some read quorum is still
available. \defword{Write fault tolerance} is defined similarly, and the
\defword{fault tolerance} of a quorum system is the minimum of its read and
write fault tolerance. For example, the read fault tolerance of $Q_{2 \times
3}$ is $1$ and the write fault tolerance is 2, so the fault tolerance is $1$.

\subsection{Load \& Capacity}
A protocol uses a \defword{strategy} to decide which quorums to contact when
executing reads and writes~\cite{Naor98}. Formally, a strategy for a quorum
system $Q = (R, W)$ is a pair $\sigma = (\sigma_R, \sigma_W)$ where $\sigma_R:
R \to [0, 1]$ and $\sigma_W: W \to [0, 1]$ are discrete probability
distributions over the quorums of $R$ and $W$. $\sigma_R(r)$ is the probability
of choosing read quorum $r$, and $\sigma_W(w)$ is the probability of choosing
write quorum $w$.  A \defword{uniform strategy} is one where each quorum is
equally likely to be chosen (i.e.\ $\sigma_R(r) = \frac{1}{|R|}$,
$\sigma_W(w) = \frac{1}{|W|}$ for every $r$ and $w$).

For a node $x \in X$, let $\text{load}_{\sigma_R}(x)$ be the probability that
$x$ is chosen by $\sigma_R$ (i.e.\ the probability that $\sigma_R$ chooses a
read quorum that contains $x$). This is called the read load on $x$. Define
$\text{load}_{\sigma_W}(x)$, the write load, similarly. Given a workload with a
\defword{read fraction} $f_r$ of reads, the load on $x$ is the probability that
$x$ is chosen by strategy $\sigma$ and is equal to $f_r
\text{load}_{\sigma_R}(x) + (1 - f_r) \text{load}_{\sigma_W}(x)$.

The most heavily loaded node is a throughput bottleneck, and its load is what
we call the load of the strategy. The \defword{load} of a quorum system is the
load of the optimal strategy (i.e.\ the strategy that achieves the lowest
load). If a quorum system has load $L$, then the inverse of the load,
$\frac{1}{L}$, is called the \defword{capacity} of the quorum system. The
capacity of a quorum system is directly proportional to the quorum system's
maximum achievable throughput.

For example, consider a 100\% read workload (i.e.\ a workload with read
fraction $f_r = 1$) and consider again the grid quorum system $Q_{2 \times 3}$
in \figref{TwoByThreeGrid}. The optimal strategy is a uniform strategy that
selects both read quorums equally likely. Thus, the load of $Q_{2 \times 3}$ is
$\frac{1}{2}$, and its capacity is 2. If every node can process $\alpha$
commands per second, then the quorum system can process $2\alpha$ commands per
second in aggregate.
Alternatively, consider a 100\% write workload with a read fraction $f_r = 0$.
The optimal strategy is again uniform. Because there are three write quorums,
the load is $\frac{1}{3}$, and the capacity is $3$. The quorum system can
process $3\alpha$ commands per second under this workload.
Finally, with $f_r = \frac{1}{2}$ (i.e.\ a workload with 50\% reads and 50\%
writes), the quorum system's capacity is $\frac{12}{5}$.
}
{\section{Practical Refinements in \library}\seclabel{PracticalRefinements}
In this section, we augment the theory of read-write quorum systems with a
number of practical considerations and demonstrate their use in our Python
library \library.

\subsection{Quorum Systems, Capacity, Fault Tolerance}
\library{} allows users to form arbitrary read-write quorum systems and compute
their capacity and fault tolerance. For example, in \figref{Basics}, we
construct and analyze the majority quorum system on nodes $\set{a, b, c}$.
As in \secref{QuorumSystemsDefinition}, read-write quorum systems are
constructed from a set of read or write quorums expressed as a boolean
expression over the set of nodes.

\begin{figure}[ht]
\begin{verbatim}
a, b, c = Node('a'), Node('b'), Node('c')
majority = QuorumSystem(reads=a*b + b*c + a*c)
print(majority.fault_tolerance())         # 1
print(majority.load(read_fraction=1))     # 2/3
print(majority.capacity(read_fraction=1)) # 3/2
\end{verbatim}
\caption{Quorum systems, capacity, and fault tolerance.}\figlabel{Basics}
\end{figure}

\newcommand{\dual}[1]{\text{dual}(#1)}
Note that the user only has to specify one set of quorums rather than both
because we automatically construct the optimal set of complementary quorums
using the existing body of literature that relates read-write quorum systems to
monotone boolean functions~\cite{Ibaraki93}.  Specifically, given a boolean
expression $e$, the \defword{dual} of $e$, denoted $\dual{e}$ is the expression
formed by swapping logical and ($\land$) with logical or ($\lor$) in $e$. For
example, $\dual{ab} = a + b$, $\dual{a + b} = ab$, and $\dual{a(b+c) + de} =
(a+bc)(d+e)$. As described in~\cite{Ibaraki93}, given a boolean expression
$e_R$ representing a set of read quorums over a set $X$, the optimal set of
complementary write quorums is $e_W = \dual{e_R}$. Similarly, given an
expression $e_W$ representing a set of write quorums, the optimal set of
complementary read quorums is $e_R = \dual{e_W}$. This is how \library{}
computes write quorums when only a set of read quorums is given (and vice
versa).

\newcommand{\opt}[1]{#1^*}

\library{} computes the load of a quorum system using linear
programming~\cite{Naor98}. Specifically, given a read-write quorum system $Q =
(R, W)$ over a set $X$ with read fraction $f_r$, we introduce a load variable
$L$, a variable $p_r$ for every $r \in R$, and a variable $p_w$ for every $w
\in W$. The linear program computes the optimal strategy $\opt{\sigma} =
(\opt{\sigma}_R, \opt{\sigma}_W)$. $L$ represents the load of $\opt{\sigma}$,
$p_r$ represents $\opt{\sigma}_R(r)$, and $p_w$ represents $\opt{\sigma}_W(w)$.
The linear program minimizes $L$ with the following constraints. First, $0 \leq
p_r, p_w \leq 1$ for every $p_r$ and $p_w$. Second, $\sum_{r \in R} p_r = 1$
and $\sum_{w \in W} p_w = 1$. These two constraints ensure that
strategies $\opt{\sigma}_R$ and $\opt{\sigma}_W$ are valid probability
distributions. Third, for every $x \in X$,
\[
  f_r \left( \sum_{\setst{r \in R}{x \in r}} p_r \right) +
  (1 - f_r) \left ( \sum_{\setst{w \in W}{x \in w}} p_w \right) \leq L
\]
This constraint ensures that the load on node $x$ is less than or equal to $L$.

\library{} computes the fault tolerance of a quorum system using integer
programming. First, we form an integer program to compute read fault tolerance.
We introduce a variable $v_x \in \set{0, 1}$ for every $x \in X$. Intuitively,
if $v_x = 1$, it means node $x$ has failed, and if $v_x = 0$, it means node $x$
is alive. We minimize $\sum_{x \in X} v_x$ with the constraint that for every
$r \in R$, $\sum_{x \in r} v_x \geq 1$. By minimizing $\sum_{x \in X} v_x$, we
try to fail as few nodes as possible. The constraint $\sum_{x \in r} v_x \geq
1$ ensures that at least one node from $r$ has failed. We then compute the read
fault tolerance as $f = (\sum_{x \in X} v_x) - 1$. $f+1$ is the minimum number
of nodes we can fail to eliminate all read quorums, so the quorum system can
tolerate as many as $f$ failures. We solve for the write fault tolerance in
the same way. The fault tolerance is the minimum of the read and write fault
tolerance.

\subsection{Heterogeneous Nodes}
Quorum system theory implicitly assumes that all nodes are equal. In reality,
nodes are often heterogeneous. Some are fast, and some are slow. Moreover,
nodes can often process more reads per second than writes per second. We revise
the theory by associating every node $x$ with its read and write capacity, i.e.\
the maximum number of reads and writes the node can process per second.
We redefine the read load imposed by a strategy $\sigma = (\sigma_R, \sigma_W)$
on a node $x$ as the probability that $\sigma_R$ chooses $x$ divided by the
read capacity of $x$. We redefine the write load similarly. By normalizing a
node's load with its capacity, we get a more intuitive definition
of a quorum system's capacity. Now, the capacity of a quorum system is the
maximum throughput that it can support.

\library{} allows users to annotate nodes with read and write capacities. For
example, in \figref{Heterogeneous}, we construct a 2 by 2 grid quorum system
where nodes $a$ and $b$ can process 100 writes per second, but nodes $c$ and
$d$ can only process 50 writes per second. We also specify that every node can
process reads twice as fast as writes.
With a read fraction of 1, the quorum system has a capacity of 300 commands per
second using a strategy that picks the read quorum $\set{a, b}$ twice as often
as the read quorum $\set{c, d}$. As we decrease the fraction of reads, the
capacity decreases since the nodes process reads faster than writes.

\begin{figure}[ht]
\begin{verbatim}
a = Node('a', write_cap=100, read_cap=200)
b = Node('b', write_cap=100, read_cap=200)
c = Node('c', write_cap=50, read_cap=100)
d = Node('d', write_cap=50, read_cap=100)
grid = QuorumSystem(reads=a*b + c*d)
print(grid.capacity(read_fraction=1))   # 300
print(grid.capacity(read_fraction=0.5)) # 200
print(grid.capacity(read_fraction=0))   # 100
\end{verbatim}
\caption{Heterogeneous nodes with different capacities.}\figlabel{Heterogeneous}
\end{figure}

\newcommand{\rcap}[1]{\text{cap}_R(#1)}
\newcommand{\wcap}[1]{\text{cap}_W(#1)}
To compute the load and capacity of a read-write quorum systems with different
read and write capacities, \library{} modifies its linear program by
normalizing every node's load by its capacity. Specifically, for every node $x
\in X$, it uses the following constraint where $\rcap{x}$ and $\wcap{x}$ are
the read and write capacities of node $x$:
\[
  \left(\frac{f_r}{\rcap{x}} \sum_{\setst{r \in R}{x \in r}} p_r\right) +
  \left(\frac{1 - f_r}{\wcap{x}} \sum_{\setst{w \in W}{x \in w}} p_w\right)
  \leq L
\]

\subsection{Workload Distributions}
Capacity is defined with respect to a fixed read and write fraction, but in
reality, workloads skew. To accommodate workload skew, we consider a discrete
probability distribution over a set of read fractions and redefine the capacity
of a quorum system to be the capacity of the strategy $\sigma$ that maximizes
the expected capacity with respect to the distribution.
For example, in \figref{ReadFractionDistribution}, we construct the quorum
system with read quorums $ac + bd$, and we define a workload that has 0\% reads
$\frac{10}{18}$th of the time, 25\% reads $\frac{4}{18}$th of the time, and so
on. In \figref{ReadFractionDistribution}, we see the optimal strategy $\sigma$
has an expected capacity of 159 commands per second.

\begin{figure}[ht]
\begin{verbatim}
grid = QuorumSystem(reads=a*c + b*d)
fr = {0.00: 10/18, 0.25: 4/18, 0.50: 2/18,
      0.75:  1/18, 1.00: 1/18}
sigma = grid.strategy(read_fraction=fr)
print(sigma.capacity(read_fraction=fr)) # 159
\end{verbatim}
\caption{A distribution of read fractions.}\figlabel{ReadFractionDistribution}
\end{figure}

In \figref{WorkloadDistribution}, we plot strategy $\sigma$'s capacity as a
function of read fraction. We also plot the capacities of strategies
$\sigma_{0.0}$, $\sigma_{0.25}$, $\sigma_{0.50}$, $\sigma_{0.75}$, and
$\sigma_{1.0}$  where $\sigma_{f_r}$ is the strategy optimized for a fixed
workload with a read fraction of $f_r$. We see that strategy $\sigma$ does not
always achieve the maximum capacity for any \emph{individual} read fraction,
but it achieves the best expected capacity across the distribution.

\begin{figure}[ht]
  \centering
  \includegraphics[]{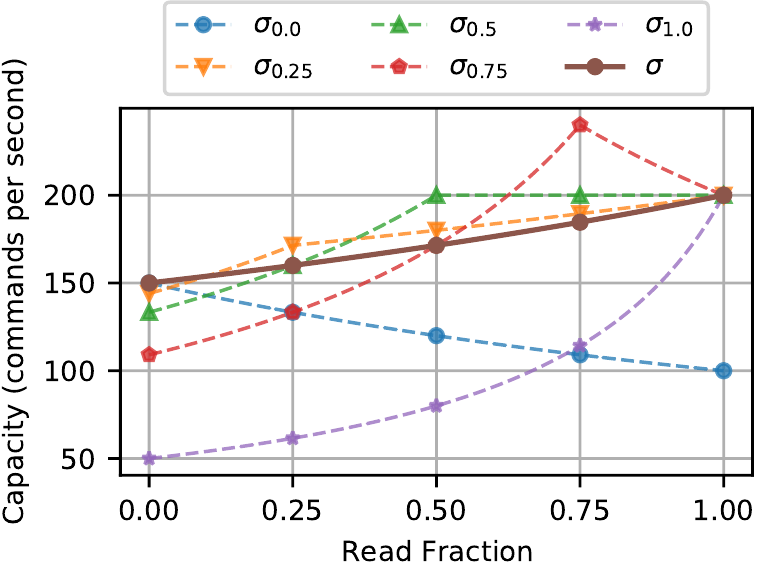}
  \caption{Strategy capacities with respect to read fraction}%
  \figlabel{WorkloadDistribution}
\end{figure}

Note that strategy $\sigma$ performs well across all workloads drawn from the
distribution without having to know the current read fraction. Alternatively,
if we are able to monitor the workload and deduce the current read fraction, we
can pre-compute a set of strategies that are optimized for various read
fractions and dynamically select the one that is best for the current workload.

To compute the load and capacity of a read-write quorum system with a
distribution of read fractions, \library{} again modifies its linear program.
Rather than minimizing a single load variable $L$, we have one load variable
$L_{f_r}$ for every possible value of $f_r$ and minimize their sum, weighted
according to their distribution. For every node $x \in X$ and every value of
read fraction $f_r$, the linear program has the constraint:
\[
  \left(\frac{f_r}{\rcap{x}} \sum_{\setst{r \in R}{x \in r}} p_r\right) +
  \left(\frac{1 - f_r}{\wcap{x}} \sum_{\setst{w \in W}{x \in w}} p_w\right)
  \leq L_{f_r}
\]

\subsection{$f$-resilient Strategies}
Many protocols that deploy read-write quorum systems actually contact more
nodes than is strictly necessary when executing a read or a write. Rather than
contacting a quorum to perform a read or write, these protocols contact
\emph{every} node. Contacting every node leads to suboptimal capacity, but it
is less sensitive to stragglers and node failures. For example, if we contact
only a quorum of nodes and one of the nodes in the quorum fails, then we have
to detect the failure and contact a different quorum. This can be slow and
costly. Typically, industry practitioners have chosen between these two
extremes: either send messages to \emph{every node} or send messages to the
\emph{bare minimum number of nodes} (i.e.\ a quorum)~\cite{shi2016cheap,
marandi2010ring, lamport2004cheap, junqueira2011zab, lakshman2010cassandra,
burrows2006chubby}. We introduce the notion of $f$-resilient strategies to show
that this is not a binary decision, but rather a continuous trade-off.

Given a quorum system $(R, W)$, we say a read quorum $r \in R$ is $f$-resilient
for some integer $f$ if despite removing any $f$ nodes from $r$, $r$ is still a
read quorum. We define $f$-resilience for write quorums similarly. We say a
strategy $\sigma$ is \defword{$f$-resilient} if it only selects $f$-resilient
read and write quorums. An $f$-resilient strategy can tolerate any $f$ failures
or stragglers. The value of $f$ captures the continuous trade-off between
capacity and resilience. As we increase $f$, we decrease capacity but increase
resilience.

\library{} allows users to compute optimal $f$-resilient strategies and their
corresponding capacities. For example, in \figref{FResilient}, we compute the
$f$-resilient capacity of a grid quorum system for $f=0$ and $f=1$.  Its
$0$-resilient capacity is $300$, but its $1$-resilient capacity is only 100.
We then compute the $f$-resilient capacities for the ``read 2, write 3''
quorum system. For this quorum system, every set of two nodes is a read quorum,
and every set of three nodes is a write quorum. This quorum system has the same
$0$-resilient capacity as the grid but a higher $1$-resilient capacity,
showing that some quorum systems are naturally more resilient than others.

\begin{figure}[ht]
\begin{verbatim}
grid = QuorumSystem(reads=a*b + c*d)
print(grid.capacity(read_fraction=1, f=0)) # 300
print(grid.capacity(read_fraction=1, f=1)) # 100
read2 = QuorumSystem(reads=choose(2, [a,b,c,d]))
print(read2.capacity(read_fraction=1, f=0)) # 300
print(read2.capacity(read_fraction=1, f=1)) # 200
\end{verbatim}
\caption{$0$-resilient and $1$-resilient strategies.}\figlabel{FResilient}
\end{figure}

\library{} computes $f$-resilient quorums using a brute-force backtracking
algorithm with pruning. Given a set of nodes $X$, \library{} enumerates every
subset of $X$ and checks whether it is an $f$-resilient quorum. However, once
an $f$-resilient quorum is found, all supersets of the quorum are pruned from
consideration.

\subsection{Latency and Network Load}
Quorum system theory focuses on capacity and fault tolerance. We introduce two
new practically important metrics. First, we introduce \defword{latency}. We
associate every node with a latency that represents the time required to
contact the node. The latency of a quorum $q$ is the time required to form a
quorum of responses after contacting the nodes in $q$. The latency of a
strategy is the expected latency of the quorums that it selects. The lower the
latency, the better. Note that if a quorum is $f$-resilient, we only need to
collect responses from at most all but $f$ of the nodes in order to form a
quorum, so the latency of a quorum can be less than the latency required to
hear back from \emph{every} node in the quorum.

Second, we introduce \defword{network load}. When a protocol executes a read or
write, it sends messages over the network to every node in a quorum, so as the
sizes of quorums increase, the number of network messages increases. The
network load of a strategy is the expected size of the quorums it chooses. The
lower the network load, the better.

In isolation, optimizing for latency or network load is trivial, but balancing
capacity, fault tolerance, latency, and network load simultaneously is very
complex. \library{} allows users to find strategies that are optimal with
respect to capacity, latency, or network load with constraints on the other
metrics. For example, in \figref{MultipleObjectives}, we specify the latencies
of the nodes in our 2 by 2 grid quorum system and then find the latency optimal
strategy with a capacity no less than 150 and with a network load of at most 2.
The optimal strategy has a latency of 3 seconds.

\begin{figure}[ht]
\small
\begin{verbatim}
a = Node('a', write_cap=100, read_cap=200, latency=4)
b = Node('b', write_cap=100, read_cap=200, latency=4)
c = Node('c', write_cap=50, read_cap=100, latency=1)
d = Node('d', write_cap=50, read_cap=100, latency=1)
grid = QuorumSystem(reads=a*b + c*d)
sigma = grid.strategy(read_fraction = 1,
                      optimize = 'latency',
                      capacity_limit = 150,
                      network_limit = 2)
print(sigma.latency(read_fraction=1)) # 3 seconds
\end{verbatim}
\caption{%
  Finding a latency-optimal strategy with capacity and network load
  constraints.
}\figlabel{MultipleObjectives}
\end{figure}

\newcommand{\latency}[1]{\text{latency}(#1)}
\library{} again uses linear programming to optimize latency and network load.
The latency of a quorum system is computed as follows where $\latency{r}$ and
$\latency{w}$ are the latencies of read quorum $r$ and write quorum $w$:
\[
  f_r \left( \sum_{r \in R} p_r \cdot \latency{r} \right) +
  (1 - f_r) \left( \sum_{w \in W} p_w \cdot \latency{w} \right)
\]
The network load is computed as
\[
  f_r \left( \sum_{r \in R} p_r \cdot |r| \right) +
  (1 - f_r) \left( \sum_{w \in W} p_w \cdot |w| \right)
\]

Note that in reality, the relationships between load, latency, and network load
are complex. For example, as the load on a node increases, the latencies of the
requests sent to it increase. Moreover, the clients that communicate with the
nodes in a quorum system may experience different latencies based on where they
are physically located. We leave these complexities to future work.

\subsection{Quorum System Search}
Thus far, we have demonstrated how \library{} makes it easy to model, analyze,
and optimize a \emph{specific hand-chosen} quorum system. \library{} also
implements a heuristic based search procedure to find good quorum systems. For
example, in \figref{Search}, we search for a quorum system over the nodes
$\set{a, b, c, d}$ optimized for latency with a capacity of at least 150 and a
network load of at most 2. The search procedure returns the quorum system with
read quorums $a+b+c+d$ and write quorums $abcd$, and with the read strategy
that picks $c$ one third of the time and $d$ two thirds of the time.

\begin{figure}[ht]
\begin{verbatim}
qs, sigma = search(nodes = [a, b, c, d],
                   read_fraction = 1,
                   optimize = 'latency',
                   capacity_limit = 150,
                   network_limit = 2)
print(qs) # reads=a+b+c+d, writes=a*b*c*d
print(sigma) # c: 1/3, d: 2/3
print(sigma.latency(read_fraction=1)) # 1 second
print(sigma.capacity(read_fraction=1)) # 150
\end{verbatim}
\caption{Searching the space of quorum systems.}\figlabel{Search}
\end{figure}

\newcommand{\echoose}[1]{\text{choose}(#1)}
Given a list of expressions $\bar{e} = e_1, \ldots, e_n$, let $\echoose{k;
\bar{e}}$ be the disjunction of the conjunction of every set of $k$
expressions in $\bar{e}$. For example, $\echoose{2; a, b, c} = ab + ac + bc$,
and $\echoose{1; a, b, c} = a + b + c$. Given a boolean expression $e$
representing a set of quorums, we say $e$ is \defword{duplicate free} if $e$
can be expressed using logical or, logical and, and choose with every variable
in $e$ appearing exactly once. For example $a + bc$ is duplicate free. $ab + ac
= a(b + c)$ is duplicate free. $ab + ac + bc = \echoose{2; a, b, c}$ is
duplicate free. $ab + ace + de + dcb$ is \emph{not} duplicate free.

Our search procedure exhaustively searches the space of all quorum systems that
have read quorums expressible by a duplicate free expression. The search
procedure heuristically explores simpler expressions first. Specifically, it
enumerates expressions in increasing order of their depth when represented as
an abstract syntax tree. Because the search space is enormous, users can
specify a timeout.
}
{\section{Case Study}\seclabel{CaseStudy}
In this section, we present a hypothetical case study that demonstrates how to
use \library{} in a realistic setting. Assume we have five nodes. Nodes $a$,
$c$, and $e$ can process 2,000 writes per second, while nodes $b$ and $d$ can
only process 1,000 writes per second. All nodes process reads twice as fast as
writes. Nodes $a$ and $b$ have a latency of 1 second; nodes $c$, $d$, and $e$
have latencies of 3, 4, and 5 seconds. We observe a workload with roughly equal
amounts of reads and writes with a slight skew towards being read heavy.  In
\figref{ModelingNodes}, we use \library{} to model the nodes and workload
distribution.

Assume we have already deployed a majority quorum system with a uniform
strategy, which has a capacity of 2,292 commands per second. We want to find a
more load optimal quorum system. We consider three candidates. The first is the
majority quorum system. The second is a staggered grid quorum system,
illustrated in \figref{StaggeredGrid}. The third is a quorum system based on
paths through a two-dimensional grid illustrated in \figref{Paths}. This quorum
system has theoretically optimal capacity~\cite{Naor98}. In
\figref{Capacities}, we construct these three quorum systems and print their
capacities.

\begin{figure}[ht]
\small
\begin{verbatim}
a = Node('a', write_cap=2000, read_cap=4000, latency=1)
b = Node('b', write_cap=1000, read_cap=2000, latency=1)
c = Node('c', write_cap=2000, read_cap=4000, latency=3)
d = Node('d', write_cap=1000, read_cap=2000, latency=4)
e = Node('e', write_cap=2000, read_cap=4000, latency=5)
fr = {0.9:  10/470, 0.8:  20/470, 0.7: 100/470,
      0.6: 100/470, 0.5: 100/470, 0.4:  60/470,
      0.3:  30/470, 0.2:  30/470, 0.1:  20/470}
\end{verbatim}
\caption{Nodes and workload distribution.}\figlabel{ModelingNodes}
\end{figure}

\begin{figure}[ht]
  \centering

  \tikzstyle{element}=[draw, circle, inner sep=0pt, minimum width=12pt, thick]
  \tikzstyle{quorum}=[rounded corners, thick]
  \tikzstyle{readquorum}=[quorum, red]
  \tikzstyle{writequorum}=[quorum, blue]

  \begin{subfigure}[b]{0.5\columnwidth}
    \centering
    \begin{tikzpicture}
      \node[element] (a) at (0, 1) {$a$};
      \node[element] (b) at (1, 1) {$b$};
      \node[element] (d) at (0, 0) {$d$};
      \node[element] (e) at (1, 0) {$e$};
      \node[element] (f) at (2, 0) {$f$};
      \draw[readquorum] (-0.3, 1.3) rectangle (1.3, 0.7);
      \draw[readquorum] (-0.3, 0.3) rectangle (2.3, -0.3);
    \end{tikzpicture}
    \caption{Staggered grid quorum system}\figlabel{StaggeredGrid}
  \end{subfigure}%
  \begin{subfigure}[b]{0.5\columnwidth}
    \centering
    \begin{tikzpicture}[]
      \node[element] (a) at (0, 1) {$a$};
      \node[element] (b) at (1, 1) {$b$};
      \node[element] (c) at (0.5, 0.5) {$c$};
      \node[element] (d) at (0, 0) {$d$};
      \node[element] (e) at (1, 0) {$e$};
      \draw[readquorum] (-0.3, 1.3) rectangle (1.3, 0.7);
      \draw[readquorum] (-0.3, 0.3) rectangle (1.3, -0.3);
      \draw[readquorum, rotate around={-45:(0, 1)}]
        (-0.3, 1.3) rectangle (1.7, 0.7);
      \draw[readquorum, rotate around={45:(0, 0)}]
        (-0.3, 0.3) rectangle (1.7, -0.3);
    \end{tikzpicture}
    \caption{Paths quorum system}\figlabel{Paths}
  \end{subfigure}
  \caption{%
    The read quorums of the staggered grid and paths quorum systems. The
    optimal set of complementary write quorums is chosen automatically.
  }
  \figlabel{GridAndPaths}
\end{figure}
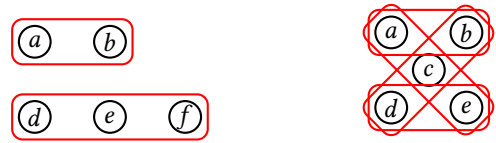

\begin{figure}[ht]
\small
\begin{verbatim}
maj = QuorumSystem(reads=majority([a, b, c, d, e]))
grid = QuorumSystem(reads=a*b + c*d*e)
paths = QuorumSystem(reads=a*b + a*c*e + d*e + d*c*b)
print(maj.capacity(reads_fraction=fr))   # 3,667
print(grid.capacity(reads_fraction=fr))  # 4,200
print(paths.capacity(reads_fraction=fr)) # 4,125
\end{verbatim}
\caption{Quorum systems and their capacities.}\figlabel{Capacities}
\end{figure}

The capacities are 3,667, 4,200, and 4,125 commands per second respectively,
making the grid quorum system the most attractive. However, the grid quorum
system is not necessarily optimal. In \figref{LoadSearched}, we perform a
search for a quorum system optimized for capacity that is tolerant to one
failure. The search takes 7 seconds on a laptop.

\begin{figure}[ht]
\begin{verbatim}
qs, sigma = search(nodes=[a, b, c, d, e],
                   fault_tolerance=1,
                   read_fraction=fr)
print(qs.capacity(read_fraction=fr)) # 5,005
\end{verbatim}
\caption{Searching for a load-optimal quorum system.}
\figlabel{LoadSearched}
\end{figure}

The search procedure finds the quorum system with read quorums $(c + bd)(a +
e)$ which has a capacity of 5,005 commands per second. This is $1.19\times$
better than the grid quorum system, and $2.18\times$ better than the majority
quorum system with a naive uniform strategy. Assume hypothetically that we
deploy this strategy to production.
Months later, we introduce a component into our system that bottlenecks our
throughput at 2,000 commands per second. Now, any capacity over 2,000 is
wasted, so we search for a quorum system optimized for latency with a
capacity of at least 2,000. We again consider our three quorum systems in
\figref{Latencies}.

\begin{figure}[ht]
\begin{verbatim}
for qs in [maj, grid, paths]:
    print(qs.latency(read_fraction=fr,
                     optimize='latency',
                     capacity_limit=2000))
\end{verbatim}
\caption{Latencies with a capacity constraint.}\figlabel{Latencies}
\end{figure}

The quorum systems have latencies of 3.24, 1.95, and 2.43 seconds respectively,
making the grid quorum system the most attractive. We again perform a search
and find the quorum system with read quorums $ab + acde + bcde$ achieves a
latency of 1.48 seconds. This is $1.32\times$ better than the grid and
$3.04\times$ better than a naive uniform strategy over a majority quorum
system. The search again completes in 7 seconds. We hypothetically deploy this
quorum system to production.
}
{\section{Lessons Learned}

\subsection{Naive Majority Quorums Underperform}
Industry practitioners often use majority quorums because they are simple and
have strong fault tolerance. Our case study shows that majority quorum systems
with uniform strategies almost always underperform more sophisticated quorum
systems in terms of capacity, latency, and network load. In
\figref{UniformMajorityThroughputs}, we plot a stacked histogram of the
throughput that every node in a majority quorum system obtains using a naive
uniform strategy, with throughput broken down by quorums. We contrast this in
\figref{OptThroughputs} with the strategy found in \figref{LoadSearched}. The
sophisticated quorum system assigns more work to machines with higher
capacities, leading to a $2.18\times$ increase in aggregate throughput.

\begin{figure}[ht]
  \centering
  \includegraphics[]{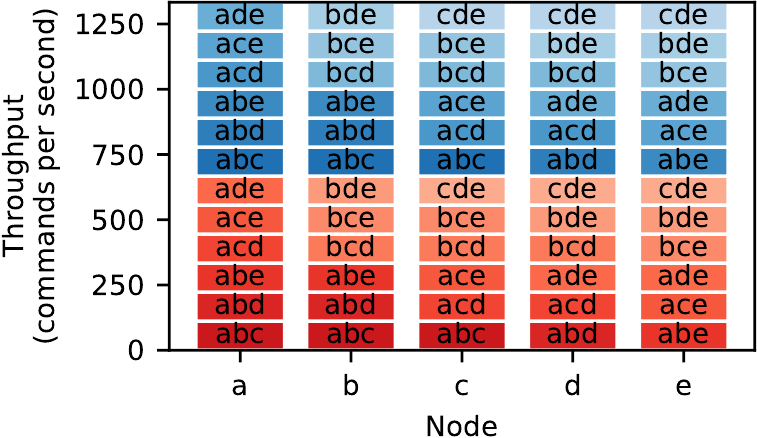}
  \caption{%
    A stacked histogram of the throughput of a simple majority quorum system
    with a naive uniform strategy. Write quorums are in blue, and read quorums
    are in red.
  }\figlabel{UniformMajorityThroughputs}
\end{figure}

\begin{figure}[ht]
  \centering
  \includegraphics[]{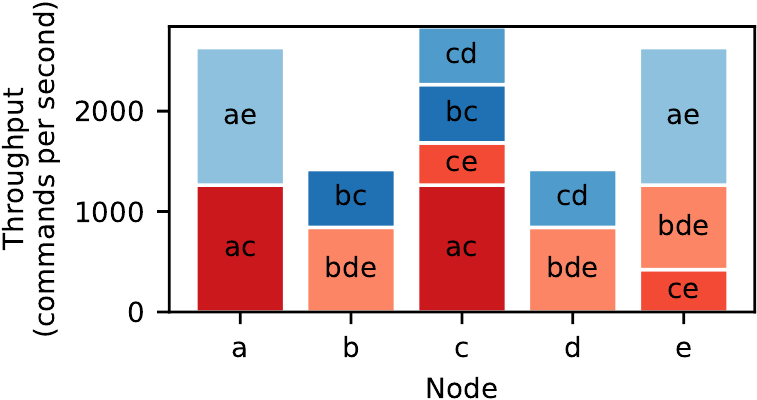}
  \caption{%
    A stacked histogram of the throughput of the quorum system found by our
    heuristic search (i.e., the quorum system with read quorums $(c+bd)(a+e)$).
  }\figlabel{OptThroughputs}
\end{figure}

\subsection{Optimal Is Not Always Optimal.}
There is a large body of research on constructing ``optimal'' quorum
systems~\cite{Garcia-Molina85, Maekawa85, Agrawal90, Kumar91, Cheung92, Naor98,
Peleg95}. For example, the paths quorum system is theoretically optimal, but in
our case study, it has lower capacity and higher latency than the simpler grid
quorum system. There are two reasons for this mismatch between theoretically
and practically optimal. First, existing quorum system theory does not account
for node heterogeneity and workload skew. Second, these quorum systems are only
optimal in the limit, as the number of nodes tends to infinity.

\subsection{The Trade-Off Space Is Complex}
Constructing a quorum system of homogeneous nodes that is optimal in the limit
for a fixed workload is difficult but doable. When nodes operate at different
speeds and workloads skew, finding an optimal quorum system that satisfies
constraints on capacity, fault tolerance, latency, and network load becomes
nearly impossible to do by hand. Moreover, small perturbations in any of these
parameters can change the landscape of the optimal quorum systems. In our case
study, for example, the search procedure finds two different quorum systems
when optimizing for load and when optimizing for latency. We believe that using
an automated assistance library like ours is the only realistic way to find
good quorum systems.
}
{\section{Conclusion}\seclabel{Conclusion}
Majority quorum systems have garnered the most attention due to their
simplicity and well-understood properties. Our tool, called \library{}, allows
engineers to find more optimal quorum systems and strategies for a given set of
constraints. With our tool, we not only show that majority quorum systems are
not necessarily the best, but we also illustrate that some quorums that
theoretically boast better performance are likely to underperform in practical
conditions.

While \library{} is easy to use, it has a few limitations that may be addressed
by future work. A more sophisticated latency model would help improve the
computation of latency-optimized strategies. We could also incorporate other
practical features such as a cost calculator and the ability to run \library{}
as a service to let applications adopt quorum systems on the fly. Additionally,
our analysis could be generalized to cover Byzantine quorum systems,
probabilistic quorum systems, and quorum systems in which every pair write
quorums have to intersect~\cite{Malkhi97, Malkhi98}.

\library{} and the associated scripts needed to reproduce this paper's
calculations are available at: \url{https://github.com/mwhittaker/quoracle}.
}

\bibliographystyle{ACM-Reference-Format}
\bibliography{references}


\begin{thebibliography}{26}


\ifx \showCODEN    \undefined \def \showCODEN     #1{\unskip}     \fi
\ifx \showDOI      \undefined \def \showDOI       #1{#1}\fi
\ifx \showISBNx    \undefined \def \showISBNx     #1{\unskip}     \fi
\ifx \showISBNxiii \undefined \def \showISBNxiii  #1{\unskip}     \fi
\ifx \showISSN     \undefined \def \showISSN      #1{\unskip}     \fi
\ifx \showLCCN     \undefined \def \showLCCN      #1{\unskip}     \fi
\ifx \shownote     \undefined \def \shownote      #1{#1}          \fi
\ifx \showarticletitle \undefined \def \showarticletitle #1{#1}   \fi
\ifx \showURL      \undefined \def \showURL       {\relax}        \fi
\providecommand\bibfield[2]{#2}
\providecommand\bibinfo[2]{#2}
\providecommand\natexlab[1]{#1}
\providecommand\showeprint[2][]{arXiv:#2}

\bibitem[\protect\citeauthoryear{Agrawal and Abbadi}{Agrawal and
  Abbadi}{1990}]%
        {Agrawal90}
\bibfield{author}{\bibinfo{person}{Divyakant Agrawal} {and}
  \bibinfo{person}{Amr~El Abbadi}.} \bibinfo{year}{1990}\natexlab{}.
\newblock \showarticletitle{The Tree Quorum Protocol: An Efficient Approach for
  Managing Replicated Data}. In \bibinfo{booktitle}{\emph{Proceedings of the
  16th International Conference on Very Large Data Bases}}
  \emph{(\bibinfo{series}{VLDB '90})}. \bibinfo{publisher}{Morgan Kaufmann
  Publishers Inc.}, \bibinfo{address}{San Francisco, CA, USA},
  \bibinfo{pages}{243--254}.
\newblock
\showISBNx{155860149X}


\bibitem[\protect\citeauthoryear{Agrawal and El~Abbadi}{Agrawal and
  El~Abbadi}{1989}]%
        {Agrawal89}
\bibfield{author}{\bibinfo{person}{Divyakant Agrawal} {and}
  \bibinfo{person}{Amr El~Abbadi}.} \bibinfo{year}{1989}\natexlab{}.
\newblock \showarticletitle{Efficient Solution to the Distributed Mutual
  Exclusion Problem}. In \bibinfo{booktitle}{\emph{Proceedings of the Eighth
  Annual ACM Symposium on Principles of Distributed Computing}} (Edmonton,
  Alberta, Canada) \emph{(\bibinfo{series}{PODC '89})}.
  \bibinfo{publisher}{Association for Computing Machinery},
  \bibinfo{address}{New York, NY, USA}, \bibinfo{pages}{193--200}.
\newblock
\showISBNx{0897913264}
\urldef\tempurl%
\url{https://doi.org/10.1145/72981.72994}
\showDOI{\tempurl}


\bibitem[\protect\citeauthoryear{Attiya, Bar-Noy, and Dolev}{Attiya
  et~al\mbox{.}}{1995}]%
        {Attiya95}
\bibfield{author}{\bibinfo{person}{Hagit Attiya}, \bibinfo{person}{Amotz
  Bar-Noy}, {and} \bibinfo{person}{Danny Dolev}.}
  \bibinfo{year}{1995}\natexlab{}.
\newblock \showarticletitle{Sharing Memory Robustly in Message-Passing
  Systems}.
\newblock \bibinfo{journal}{\emph{J. ACM}} \bibinfo{volume}{42},
  \bibinfo{number}{1} (\bibinfo{date}{Jan.} \bibinfo{year}{1995}),
  \bibinfo{pages}{124--142}.
\newblock
\showISSN{0004-5411}
\urldef\tempurl%
\url{https://doi.org/10.1145/200836.200869}
\showDOI{\tempurl}


\bibitem[\protect\citeauthoryear{Burrows}{Burrows}{2006}]%
        {burrows2006chubby}
\bibfield{author}{\bibinfo{person}{Mike Burrows}.}
  \bibinfo{year}{2006}\natexlab{}.
\newblock \showarticletitle{The Chubby lock service for loosely-coupled
  distributed systems}. In \bibinfo{booktitle}{\emph{Proceedings of the 7th
  symposium on Operating systems design and implementation}}.
  \bibinfo{pages}{335--350}.
\newblock


\bibitem[\protect\citeauthoryear{Charapko, Ailijiang, and Demirbas}{Charapko
  et~al\mbox{.}}{2019}]%
        {Charapko19}
\bibfield{author}{\bibinfo{person}{Aleksey Charapko}, \bibinfo{person}{Ailidani
  Ailijiang}, {and} \bibinfo{person}{Murat Demirbas}.}
  \bibinfo{year}{2019}\natexlab{}.
\newblock \showarticletitle{Linearizable Quorum Reads in Paxos}. In
  \bibinfo{booktitle}{\emph{Proceedings of the 11th USENIX Conference on Hot
  Topics in Storage and File Systems}} (Renton, WA, USA)
  \emph{(\bibinfo{series}{HotStorage'19})}. \bibinfo{publisher}{USENIX
  Association}.
\newblock


\bibitem[\protect\citeauthoryear{Cheung, Ahamad, and Ammar}{Cheung
  et~al\mbox{.}}{1990}]%
        {Cheung90}
\bibfield{author}{\bibinfo{person}{Shun~Yan Cheung}, \bibinfo{person}{Mustaque
  Ahamad}, {and} \bibinfo{person}{Mostafa~H. Ammar}.}
  \bibinfo{year}{1990}\natexlab{}.
\newblock \showarticletitle{Multidimensional voting: a general method for
  implementing synchronization in distributed systems}. In
  \bibinfo{booktitle}{\emph{Proceedings of the 10th International Conference on
  Distributed Computing Systems}}. \bibinfo{pages}{362--369}.
\newblock
\urldef\tempurl%
\url{https://doi.org/10.1109/ICDCS.1990.89304}
\showDOI{\tempurl}


\bibitem[\protect\citeauthoryear{Cheung, Ammar, and Ahamad}{Cheung
  et~al\mbox{.}}{1992}]%
        {Cheung92}
\bibfield{author}{\bibinfo{person}{Shun~Yan Cheung},
  \bibinfo{person}{Mostafa~H. Ammar}, {and} \bibinfo{person}{Mustaque Ahamad}.}
  \bibinfo{year}{1992}\natexlab{}.
\newblock \showarticletitle{The Grid Protocol: A High Performance Scheme for
  Maintaining Replicated Data}.
\newblock \bibinfo{journal}{\emph{IEEE Trans. on Knowl. and Data Eng.}}
  \bibinfo{volume}{4}, \bibinfo{number}{6} (\bibinfo{date}{Dec.}
  \bibinfo{year}{1992}), \bibinfo{pages}{582--592}.
\newblock
\showISSN{1041-4347}
\urldef\tempurl%
\url{https://doi.org/10.1109/69.180609}
\showDOI{\tempurl}


\bibitem[\protect\citeauthoryear{Fischer, Lynch, and Paterson}{Fischer
  et~al\mbox{.}}{1985}]%
        {Fischer85}
\bibfield{author}{\bibinfo{person}{Michael~J. Fischer},
  \bibinfo{person}{Nancy~A. Lynch}, {and} \bibinfo{person}{Michael~S.
  Paterson}.} \bibinfo{year}{1985}\natexlab{}.
\newblock \showarticletitle{Impossibility of Distributed Consensus with One
  Faulty Process}.
\newblock \bibinfo{journal}{\emph{J. ACM}} \bibinfo{volume}{32},
  \bibinfo{number}{2} (\bibinfo{date}{April} \bibinfo{year}{1985}),
  \bibinfo{pages}{374--382}.
\newblock
\showISSN{0004-5411}
\urldef\tempurl%
\url{https://doi.org/10.1145/3149.214121}
\showDOI{\tempurl}


\bibitem[\protect\citeauthoryear{Garcia-Molina and Barbara}{Garcia-Molina and
  Barbara}{1985}]%
        {Garcia-Molina85}
\bibfield{author}{\bibinfo{person}{Hector Garcia-Molina} {and}
  \bibinfo{person}{Daniel Barbara}.} \bibinfo{year}{1985}\natexlab{}.
\newblock \showarticletitle{How to Assign Votes in a Distributed System}.
\newblock \bibinfo{journal}{\emph{J. ACM}} \bibinfo{volume}{32},
  \bibinfo{number}{4} (\bibinfo{date}{Oct.} \bibinfo{year}{1985}),
  \bibinfo{pages}{841--860}.
\newblock
\showISSN{0004-5411}
\urldef\tempurl%
\url{https://doi.org/10.1145/4221.4223}
\showDOI{\tempurl}


\bibitem[\protect\citeauthoryear{Gifford}{Gifford}{1979}]%
        {Gifford79}
\bibfield{author}{\bibinfo{person}{David~K. Gifford}.}
  \bibinfo{year}{1979}\natexlab{}.
\newblock \showarticletitle{Weighted Voting for Replicated Data}. In
  \bibinfo{booktitle}{\emph{Proceedings of the Seventh ACM Symposium on
  Operating Systems Principles}} \emph{(\bibinfo{series}{SOSP '79})}.
  \bibinfo{publisher}{Association for Computing Machinery},
  \bibinfo{address}{New York, NY, USA}, \bibinfo{pages}{150--162}.
\newblock
\showISBNx{0897910095}
\urldef\tempurl%
\url{https://doi.org/10.1145/800215.806583}
\showDOI{\tempurl}


\bibitem[\protect\citeauthoryear{Herlihy}{Herlihy}{1986}]%
        {Herlihy86}
\bibfield{author}{\bibinfo{person}{Maurice Herlihy}.}
  \bibinfo{year}{1986}\natexlab{}.
\newblock \showarticletitle{A Quorum-Consensus Replication Method for Abstract
  Data Types}.
\newblock \bibinfo{journal}{\emph{ACM Trans. Comput. Syst.}}
  \bibinfo{volume}{4}, \bibinfo{number}{1} (\bibinfo{date}{Feb.}
  \bibinfo{year}{1986}), \bibinfo{pages}{32--53}.
\newblock
\showISSN{0734-2071}
\urldef\tempurl%
\url{https://doi.org/10.1145/6306.6308}
\showDOI{\tempurl}


\bibitem[\protect\citeauthoryear{Howard, Malkhi, and Spiegelman}{Howard
  et~al\mbox{.}}{2017}]%
        {Howard2017}
\bibfield{author}{\bibinfo{person}{Heidi Howard}, \bibinfo{person}{Dahlia
  Malkhi}, {and} \bibinfo{person}{Alexander Spiegelman}.}
  \bibinfo{year}{2017}\natexlab{}.
\newblock \showarticletitle{Flexible Paxos: Quorum Intersection Revisited}. In
  \bibinfo{booktitle}{\emph{20th International Conference on Principles of
  Distributed Systems (OPODIS 2016)}}. Schloss Dagstuhl-Leibniz-Zentrum fuer
  Informatik.
\newblock


\bibitem[\protect\citeauthoryear{Ibaraki and Kameda}{Ibaraki and
  Kameda}{1993}]%
        {Ibaraki93}
\bibfield{author}{\bibinfo{person}{Toshihide Ibaraki} {and}
  \bibinfo{person}{Tiko Kameda}.} \bibinfo{year}{1993}\natexlab{}.
\newblock \showarticletitle{A theory of coteries: Mutual exclusion in
  distributed systems}.
\newblock \bibinfo{journal}{\emph{IEEE Transactions on Parallel and Distributed
  Systems}} \bibinfo{volume}{4}, \bibinfo{number}{7} (\bibinfo{year}{1993}),
  \bibinfo{pages}{779--794}.
\newblock


\bibitem[\protect\citeauthoryear{Junqueira, Reed, and Serafini}{Junqueira
  et~al\mbox{.}}{2011}]%
        {junqueira2011zab}
\bibfield{author}{\bibinfo{person}{Flavio~P Junqueira},
  \bibinfo{person}{Benjamin~C Reed}, {and} \bibinfo{person}{Marco Serafini}.}
  \bibinfo{year}{2011}\natexlab{}.
\newblock \showarticletitle{Zab: High-performance broadcast for primary-backup
  systems}. In \bibinfo{booktitle}{\emph{2011 IEEE/IFIP 41st International
  Conference on Dependable Systems \& Networks (DSN)}}. IEEE,
  \bibinfo{pages}{245--256}.
\newblock


\bibitem[\protect\citeauthoryear{Kumar}{Kumar}{1991}]%
        {Kumar91}
\bibfield{author}{\bibinfo{person}{Akhil Kumar}.}
  \bibinfo{year}{1991}\natexlab{}.
\newblock \showarticletitle{Hierarchical Quorum Consensus: A New Algorithm for
  Managing Replicated Data}.
\newblock \bibinfo{journal}{\emph{IEEE Trans. Comput.}} \bibinfo{volume}{40},
  \bibinfo{number}{9} (\bibinfo{date}{Sept.} \bibinfo{year}{1991}),
  \bibinfo{pages}{996--1004}.
\newblock
\showISSN{0018-9340}
\urldef\tempurl%
\url{https://doi.org/10.1109/12.83661}
\showDOI{\tempurl}


\bibitem[\protect\citeauthoryear{Lakshman and Malik}{Lakshman and
  Malik}{2010}]%
        {lakshman2010cassandra}
\bibfield{author}{\bibinfo{person}{Avinash Lakshman} {and}
  \bibinfo{person}{Prashant Malik}.} \bibinfo{year}{2010}\natexlab{}.
\newblock \showarticletitle{Cassandra: a decentralized structured storage
  system}.
\newblock \bibinfo{journal}{\emph{ACM SIGOPS Operating Systems Review}}
  \bibinfo{volume}{44}, \bibinfo{number}{2} (\bibinfo{year}{2010}),
  \bibinfo{pages}{35--40}.
\newblock


\bibitem[\protect\citeauthoryear{Lamport}{Lamport}{1998}]%
        {Lamport98}
\bibfield{author}{\bibinfo{person}{Leslie Lamport}.}
  \bibinfo{year}{1998}\natexlab{}.
\newblock \showarticletitle{The Part-Time Parliament}.
\newblock \bibinfo{journal}{\emph{ACM Trans. Comput. Syst.}}
  \bibinfo{volume}{16}, \bibinfo{number}{2} (\bibinfo{date}{May}
  \bibinfo{year}{1998}), \bibinfo{pages}{133--169}.
\newblock
\showISSN{0734-2071}
\urldef\tempurl%
\url{https://doi.org/10.1145/279227.279229}
\showDOI{\tempurl}


\bibitem[\protect\citeauthoryear{Lamport and Massa}{Lamport and Massa}{2004}]%
        {lamport2004cheap}
\bibfield{author}{\bibinfo{person}{Leslie Lamport} {and} \bibinfo{person}{Mike
  Massa}.} \bibinfo{year}{2004}\natexlab{}.
\newblock \showarticletitle{Cheap paxos}. In
  \bibinfo{booktitle}{\emph{International Conference on Dependable Systems and
  Networks, 2004}}. IEEE, \bibinfo{pages}{307--314}.
\newblock


\bibitem[\protect\citeauthoryear{Maekawa}{Maekawa}{1985}]%
        {Maekawa85}
\bibfield{author}{\bibinfo{person}{Mamoru Maekawa}.}
  \bibinfo{year}{1985}\natexlab{}.
\newblock \showarticletitle{A square root N Algorithm for Mutual Exclusion in
  Decentralized Systems}.
\newblock \bibinfo{journal}{\emph{ACM Trans. Comput. Syst.}}
  \bibinfo{volume}{3}, \bibinfo{number}{2} (\bibinfo{date}{May}
  \bibinfo{year}{1985}), \bibinfo{pages}{145--159}.
\newblock
\showISSN{0734-2071}
\urldef\tempurl%
\url{https://doi.org/10.1145/214438.214445}
\showDOI{\tempurl}


\bibitem[\protect\citeauthoryear{Malkhi and Reiter}{Malkhi and Reiter}{1998}]%
        {Malkhi98}
\bibfield{author}{\bibinfo{person}{Dahlia Malkhi} {and}
  \bibinfo{person}{Michael Reiter}.} \bibinfo{year}{1998}\natexlab{}.
\newblock \showarticletitle{Byzantine Quorum Systems}.
\newblock \bibinfo{journal}{\emph{Distrib. Comput.}} \bibinfo{volume}{11},
  \bibinfo{number}{4} (\bibinfo{date}{Oct.} \bibinfo{year}{1998}),
  \bibinfo{pages}{203--213}.
\newblock
\showISSN{0178-2770}
\urldef\tempurl%
\url{https://doi.org/10.1007/s004460050050}
\showDOI{\tempurl}


\bibitem[\protect\citeauthoryear{Malkhi, Reiter, and Wright}{Malkhi
  et~al\mbox{.}}{1997}]%
        {Malkhi97}
\bibfield{author}{\bibinfo{person}{Dahlia Malkhi}, \bibinfo{person}{Michael
  Reiter}, {and} \bibinfo{person}{Rebecca Wright}.}
  \bibinfo{year}{1997}\natexlab{}.
\newblock \showarticletitle{Probabilistic Quorum Systems}. In
  \bibinfo{booktitle}{\emph{Proceedings of the Sixteenth Annual ACM Symposium
  on Principles of Distributed Computing}} (Santa Barbara, California, USA)
  \emph{(\bibinfo{series}{PODC '97})}. \bibinfo{publisher}{Association for
  Computing Machinery}, \bibinfo{address}{New York, NY, USA},
  \bibinfo{pages}{267--273}.
\newblock
\showISBNx{0897919521}
\urldef\tempurl%
\url{https://doi.org/10.1145/259380.259458}
\showDOI{\tempurl}


\bibitem[\protect\citeauthoryear{Marandi, Primi, Schiper, and Pedone}{Marandi
  et~al\mbox{.}}{2010}]%
        {marandi2010ring}
\bibfield{author}{\bibinfo{person}{Parisa~Jalili Marandi},
  \bibinfo{person}{Marco Primi}, \bibinfo{person}{Nicolas Schiper}, {and}
  \bibinfo{person}{Fernando Pedone}.} \bibinfo{year}{2010}\natexlab{}.
\newblock \showarticletitle{Ring Paxos: A high-throughput atomic broadcast
  protocol}. In \bibinfo{booktitle}{\emph{2010 IEEE/IFIP International
  Conference on Dependable Systems \& Networks (DSN)}}. IEEE,
  \bibinfo{pages}{527--536}.
\newblock


\bibitem[\protect\citeauthoryear{Naor and Wool}{Naor and Wool}{1998}]%
        {Naor98}
\bibfield{author}{\bibinfo{person}{Moni Naor} {and} \bibinfo{person}{Avishai
  Wool}.} \bibinfo{year}{1998}\natexlab{}.
\newblock \showarticletitle{The Load, Capacity, and Availability of Quorum
  Systems}.
\newblock \bibinfo{journal}{\emph{SIAM J. Comput.}} \bibinfo{volume}{27},
  \bibinfo{number}{2} (\bibinfo{year}{1998}).
\newblock
\showISSN{0097-5397}
\urldef\tempurl%
\url{https://doi.org/10.1137/S0097539795281232}
\showDOI{\tempurl}


\bibitem[\protect\citeauthoryear{Peleg and Wool}{Peleg and Wool}{1995}]%
        {Peleg95}
\bibfield{author}{\bibinfo{person}{David Peleg} {and} \bibinfo{person}{Avishai
  Wool}.} \bibinfo{year}{1995}\natexlab{}.
\newblock \showarticletitle{Crumbling Walls: A Class of Practical and Efficient
  Quorum Systems}. In \bibinfo{booktitle}{\emph{Proceedings of the Fourteenth
  Annual ACM Symposium on Principles of Distributed Computing}} (Ottowa,
  Ontario, Canada) \emph{(\bibinfo{series}{PODC '95})}.
  \bibinfo{publisher}{Association for Computing Machinery},
  \bibinfo{address}{New York, NY, USA}, \bibinfo{pages}{120--129}.
\newblock
\showISBNx{0897917103}
\urldef\tempurl%
\url{https://doi.org/10.1145/224964.224978}
\showDOI{\tempurl}


\bibitem[\protect\citeauthoryear{Shi and Wang}{Shi and Wang}{2016}]%
        {shi2016cheap}
\bibfield{author}{\bibinfo{person}{Rong Shi} {and} \bibinfo{person}{Yang
  Wang}.} \bibinfo{year}{2016}\natexlab{}.
\newblock \showarticletitle{Cheap and available state machine replication}. In
  \bibinfo{booktitle}{\emph{2016 {USENIX} Annual Technical Conference
  ({USENIX}{ATC} 16)}}. \bibinfo{pages}{265--279}.
\newblock


\bibitem[\protect\citeauthoryear{Thomas}{Thomas}{1979}]%
        {Thomas79}
\bibfield{author}{\bibinfo{person}{Robert~H. Thomas}.}
  \bibinfo{year}{1979}\natexlab{}.
\newblock \showarticletitle{A Majority Consensus Approach to Concurrency
  Control for Multiple Copy Databases}.
\newblock \bibinfo{journal}{\emph{ACM Trans. Database Syst.}}
  \bibinfo{volume}{4}, \bibinfo{number}{2} (\bibinfo{date}{June}
  \bibinfo{year}{1979}), \bibinfo{pages}{180--209}.
\newblock
\showISSN{0362-5915}
\urldef\tempurl%
\url{https://doi.org/10.1145/320071.320076}
\showDOI{\tempurl}


\end{thebibliography}

\end{document}